\documentclass[prl,twocolumn]{revtex4}
\usepackage{graphicx}

\begin{document}
\title{Atom-molecule Rabi oscillations in a Mott insulator}
\author{N. Syassen, D.~M. Bauer, M. Lettner, D. Dietze, T. Volz, S. D\"{u}rr, and G. Rempe}
\affiliation{Max-Planck-Institut f{\"u}r Quantenoptik, Hans-Kopfermann-Stra{\ss}e 1, 85748 Garching, Germany}

\hyphenation{Fesh-bach mol-e-cule mol-e-cules mi-cro-wave res-o-nance res-o-nances}

\begin{abstract}
We observe large-amplitude Rabi oscillations between an atomic and a molecular state near a Feshbach resonance. The experiment uses $^{87}$Rb in an optical lattice and a Feshbach resonance near 414~G. The frequency and amplitude of the oscillations depend on magnetic field in a way that is well described by a two-level model. The observed density dependence of the oscillation frequency agrees with the theoretical expectation. We confirmed that the state produced after a half-cycle contains exactly one molecule at each lattice site. In addition, we show that for energies in a gap of the lattice band structure, the molecules cannot dissociate.
\end{abstract}


\maketitle

In recent years, the field of ultracold molecular gases has undergone rapid progress. These systems have the potential for a variety of interesting applications, {\em e.g.}, in precision measurements or in quantum simulations. As precision measurements are usually best performed in terms of frequency measurements, atom-molecule Rabi oscillations open up promising opportunities in this direction. Specifically, it was proposed recently that measurements of atomic scattering properties with moderate accuracy might be used to perform sensitive tests for drifts of fundamental constants \cite{chin:06}. Most of the work on molecule association so far was based on adiabatically ramping a magnetic field across a Feshbach resonance \cite{regal:03,herbig:03,duerr:04,strecker:03,cubizolles:03,xu:03,koehler:06}. Unlike adiabatic ramps, time-resolved Rabi oscillations allow for full control over the final superposition state produced. The pioneering experiment on molecule association recorded Ramsey oscillations between the atomic and the molecular state, but this experiment observed Rabi oscillations only ``over a very limited range'' of the magnetic field \cite{donley:02}. In a subsequent experiment, atom-molecule Rabi oscillations with 6\% amplitude were induced using a radio-frequency field \cite{thompson:05a}. Atom-molecule Rabi oscillations were also reported in a photoassociation experiment \cite{ryu:cond-mat/0508201}. In addition, oscillations between two bound molecular states were observed \cite{mark:arXiv:0704.0653}.

Here we report the experimental observation of time-resolved Rabi oscillations between the atomic and the molecular state with large amplitude. The oscillations are weakly damped and the data show oscillations up to the $29^{\rm th}$ cycle. The observation of Rabi oscillations requires a pulse shape that is rectangular, or at least strongly diabatic. In free space, such pulses populate \cite{donley:02} the continuum of above-threshold entrance-channel states thus leading to oscillations between many levels, typically with small molecular amplitude. We avoid this by working in a deep three-dimensional optical lattice, where the entrance-channel states are discrete. For weak enough coupling, the coupling of the molecular state to only one entrance-channel state is noticeable. In addition, the lattice isolates the molecules from each other, thus suppressing loss due to inelastic collisions \cite{thalhammer:06}. Our experiment starts from an atomic Mott insulator \cite{greiner:02} prepared such that the central region of the cloud contains exactly two atoms at each lattice site. The quantum state reached after a half-cycle of the atom-molecule Rabi oscillation therefore contains exactly one molecule at each lattice site in this central region. We previously prepared the same state using an adiabatic ramp of the magnetic field \cite{volz:06,duerr:06:short}. In addition, we show that confinement-induced molecules exist not only below the lowest band of the lattice \cite{moritz:05}, but also in band gaps.

We use an optical lattice that is deep enough that tunneling is negligible. Here, each lattice site represents a simple harmonic trap with angular frequency $\omega_{ho}=k\sqrt{2V_0/m}$, where $2\pi/k=830.44$~nm is the wavelength of the lattice light, $m$ the mass of an atom, and $V_0$ the lattice depth seen by an atom \cite{volz:06}. In contrast to the free-space case, the closed-channel molecular state $|\psi_m\rangle$ is coupled to only one discrete state, namely the motional ground state $|\psi_a\rangle$ of two entrance-channel atoms at one lattice site. The matrix element $H_{am}=\langle \psi_a | H | \psi_m \rangle$ of the Hamiltonian $H$ is \cite{atom-molecule-paper:EPAPS,atom-molecule-paper:julienne}
\begin{eqnarray}
\label{eq-matrix-element}
H_{am} =
\left[
\frac{4\pi \hbar^2 a_{bg} \Delta\mu \Delta B}{m \left( \sqrt{2\pi} \; a_{ho}\right)^3}
\left( 1+ 0.490 \frac{a_{bg}}{a_{ho}} \right)
\right]^{1/2}
,
\end{eqnarray}
where $a_{ho} = \sqrt{\hbar/m\omega_{ho}}$ is the harmonic oscillator length, $a_{bg}$ the background scattering length, $\Delta B$ the width of the Feshbach resonance, and $\Delta \mu$ the difference between the magnetic moments of an entrance-channel atom pair and a closed-channel molecule. At resonance, Rabi oscillations between the two states are expected to occur with angular frequency $\Omega_{\rm res} =  2 H_{am} /\hbar$.

\begin{figure*}[t!]
\includegraphics[scale=1]{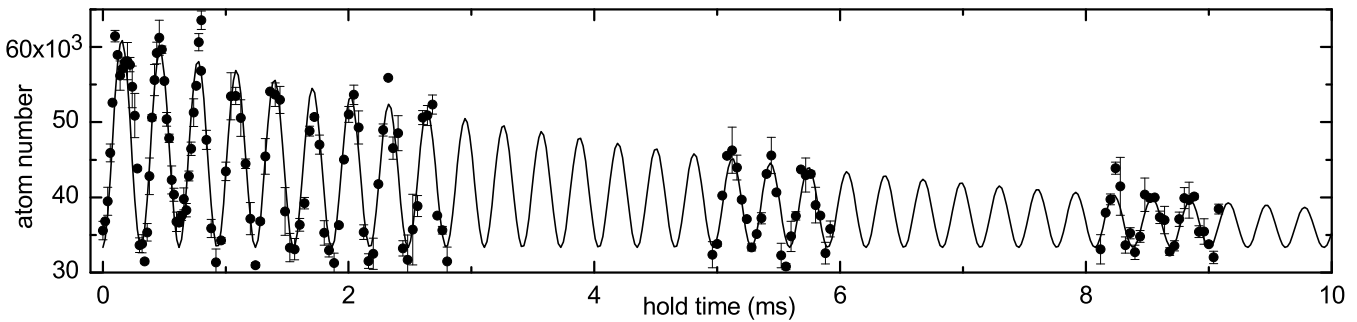}
\caption{\label{fig-oscillation}
Time-resolved Rabi oscillations between the atomic and the molecular state. The experimental data ($\bullet$) show the number of entrance-channel atoms. The line shows a fit of Eq.~(\ref{eq-damped-sine}) that yields $\Omega_{\rm Rabi}=2\pi\times 3.221(2)$~kHz and $\tau=5.9(4)$~ms.
}
\end{figure*}

In addition to the coupling in Eq.~(\ref{eq-matrix-element}), $|\psi_m\rangle$ can couple to excited $s$-wave trap states of the relative motion of two entrance-channel atoms \cite{koehler:06}. This coupling between many states can lead to quite complex dynamics. In order to avoid this, we need a Feshbach resonance with $\Omega_{\rm res} \ll \omega_{ho}$. For very small $\Omega_{\rm res}$, the magnetic field noise $\delta B_{\rm rms}$ \cite{duerr:04a} becomes an issue, resulting in the condition $\delta B_{\rm rms} \Delta\mu /\hbar \ll \Omega_{\rm res} \ll \omega_{ho}$. In the experiment, we choose a Feshbach resonance in $^{87}$Rb near 414~G with both incoming atoms in the hyperfine state $|F=1,m_F=0\rangle$ \cite{marte:02}. A coupled-channels calculation \cite{kokkelmans:pers} predicts $a_{bg}=100.8$ Bohr radii and $\Delta B=18$~mG. The Breit-Rabi formula predicts $\Delta\mu=2\pi\hbar\times 111$~kHz/G. This is an unusually small value that helps reducing $\Omega_{\rm res}$ as well as the sensitivity to magnetic field noise.

Our experiment starts with the preparation of an ultracold gas of $^{87}$Rb atoms in the hyperfine state $|1,-1\rangle$ in a magnetic trap. When the gas is just above the critical temperature for Bose-Einstein condensation (BEC), it is transferred into an optical dipole trap, where a magnetic field of $\sim 1$~G is applied to preserve the spin polarization of the atoms. Next, a magnetic field of $\sim 412$~G is switched on rapidly. We deliberately create an angle between the two fields, so that the sudden turn-on populates all $F=1$ Zeeman states. We then drive evaporative cooling by lowering the dipole trap potential. Due to the presence of a slight magnetic-field gradient, this preferentially removes atoms in states $|1,-1\rangle$ and $|1,1\rangle$. We obtain an almost pure BEC with more than 90\% of the population in the state $|1,0\rangle$. Next, an optical lattice is slowly ramped up. The experimental parameters are chosen in such a way that the central region of the resulting Mott insulator contains exactly two atoms at each lattice site \cite{volz:06}. Unless otherwise noted, the lattice depth seen by an atom is $V_0=15\,E_r$, where $E_r=\hbar^2k^2/2m$ is the atomic recoil energy.

Subsequently, the magnetic field is jumped to a value very close to the 414-G Feshbach resonance. In response to the step in the externally applied field, eddy currents build up. We therefore use two subsequent steps of the magnetic field. The first step begins 2.4~G below the Feshbach resonance and ends typically 50~mG away from the Feshbach resonance, where mixing between states $|\psi_a\rangle$ and $|\psi_m\rangle$ is negligible. 250~$\mu$s later the eddy currents have fully settled and the second step is applied. The height of the second step is small enough that eddy currents have negligible effect. After this step, we hold the magnetic field for a variable time. Finally, magnetic field, lattice, and dipole trap are abruptly switched off and after 4~ms of free flight an absorption image is taken. The imaging light is resonant with an atomic transition so that molecules remain invisible.

The number of atoms as a function of hold time right at the Feshbach resonance $B_{\rm res}$ is shown in Fig.~\ref{fig-oscillation}. The experimental data clearly show atom-molecule Rabi oscillations up to the $29^{\rm th}$ cycle. The data show damping in a way that the minimum atom number is essentially unchanged. This suggests that the decay is due to loss of population, as opposed to dephasing which would lead to damping towards the mean atom number. We therefore fit \cite{atom-molecule-paper:eddy}
\begin{eqnarray}
\label{eq-damped-sine}
N(t)=N_1 + N_2 e^{-t/\tau} \frac{1- \cos(\Omega_{\rm Rabi} t )}2
\end{eqnarray}
to the data. Both, $|\psi_a\rangle$ and $|\psi_m\rangle$ can decay into lower-lying open two-atom channels. We measured the decay of population in state $|\psi_a\rangle$ 2.4~G below the Feshbach resonance and obtained a decay rate of less than 1~Hz. We therefore attribute the decay observed in Fig.\ref{fig-oscillation} fully to the state $|\psi_m\rangle$. During a Rabi cycle, half of the time on average is spent in this state. Hence, the decay rate $\Gamma$ of population in $|\psi_m\rangle$ can be extracted from the fit of Eq.~(\ref{eq-damped-sine}) yielding $\Gamma=2/\tau=0.34(2)$~kHz.

The fraction of the population that participates in the oscillation at short time is $N_2/(N_1+N_2)=0.46(1)$. This value reflects the fraction of lattice sites that are initially occupied by two atoms \cite{volz:06}. The conversion efficiency at these sites is nearly 100\%. We repeated the measurements of Ref.~\cite{volz:06} in order to verify that the state produced here really is a quantum state in which the central region of the cloud contains exactly one molecule at each lattice site.

\begin{figure}[t!]
\includegraphics[scale=.5]{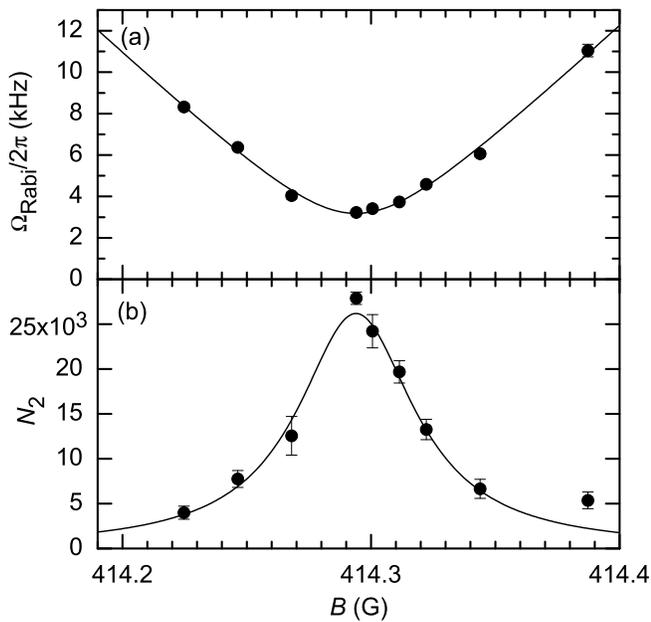}
\caption{\label{fig-freq-ampl}
Magnetic-field dependence of (a) the frequency and (b) the amplitude of the Rabi oscillations. The lines show fits to the experimental data ($\bullet$).
}
\end{figure}

The frequency and amplitude of the Rabi oscillations depend on the magnetic-field value during the hold time after the second step. This dependence is shown in Fig.~\ref{fig-freq-ampl}. As in any two-level system, the Rabi frequency is expected to follow a hyperbola
\begin{eqnarray}
\label{eq-hyperbola}
\Omega_{\rm Rabi}(B) = \sqrt{ \Omega_{\rm res}^2 + [(B-B_{\rm res}) \Delta \mu/\hbar]^2}
\; ,
\end{eqnarray}
where $B_{\rm res}$ is the resonance position which depends on the lattice depth $V_0$, as discussed further below. A fit to the data is shown in Fig.~\ref{fig-freq-ampl}(a). The best-fit values are $\Omega_{\rm res}=2\pi\times 3.2(1)$~kHz and $\Delta \mu = 2\pi\hbar \times 112(2)$~kHz/G, in good agreement with the result of Fig.~\ref{fig-oscillation} and the theoretical prediction, respectively.

The amplitude of the Rabi oscillation is shown in Fig.~\ref{fig-freq-ampl}(b) as a function of magnetic field. This amplitude follows a Lorentzian $N_2(B) =N_{\rm res} \Omega_{\rm res}^2 / \Omega_{\rm Rabi}^2(B)$ which is shown in Fig.~\ref{fig-freq-ampl}(b). We use only $N_{\rm res}$ as a free fit parameter and copy the values of the other parameters from the fit to Fig.~\ref{fig-freq-ampl}(a).

\begin{figure}[t!]
\includegraphics[scale=.5]{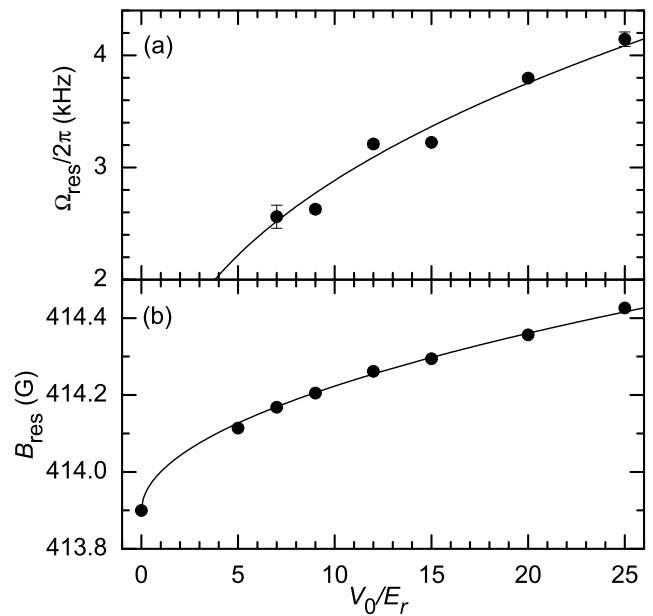}
\caption{\label{fig-density}
Dependence of (a) the on-resonance Rabi frequency and (b) the Feshbach resonance position on lattice depth. The lines show fits to the data ($\bullet$). 
}
\end{figure}

$\Omega_{\rm res}$ depends on the atomic density in the entrance-channel state. In Eq.~(\ref{eq-matrix-element}) the corresponding effective volume is $(\sqrt{2\pi}a_{ho})^3/(1+ 0.49 a_{bg}/a_{ho})$. We varied the lattice depth $V_0$ in order to verify this density dependence. Results are shown in Fig.~\ref{fig-density}(a). The line shows a fit of Eq.~(\ref{eq-matrix-element}) to the data, where the only free fit parameter is the overall amplitude. As $\Delta\mu$ and $a_{bg}$ can typically be predicted much more accurately than $\Delta B$, we use this fit to determine $\Delta B=15(1)$~mG which agrees fairly well with theory. The dominant contribution to the error in $\Delta B$ comes from the calibration of $V_0$, which we perform in terms of a frequency measurement \cite{duerr:06:short}. We estimate the relative error in $V_0$ to be 10\%. This determination of $\Delta B$ is independent of the atom-number calibration, because only lattice sites with exactly two atoms contribute to the oscillation.

The measurements of $\Omega_{\rm res}$ in Fig.~\ref{fig-density}(a) rely on a measurement of $B_{\rm res}$ as a function of lattice depth $V_0$. Results of this measurement are shown in Fig.~\ref{fig-density}(b). Based on the zero-point energy of the three-dimensional harmonic oscillator for the relative motion of the two atoms, one expects \cite{moritz:05} $B_{\rm res} = B_0 + 3\hbar \omega_{ho}/2\Delta\mu$, where $B_0$ is the value at $V_0=0$. The background scattering length $a_{bg}$ causes a correction \cite{busch:98} yielding
\begin{eqnarray}
\label{eq-B-res-shift}
B_{\rm res} 
= B_0 + \frac{\hbar \omega_{ho}}{\Delta\mu} \left(\frac32 + \sqrt{\frac2\pi} \; \frac{a_{bg}}{a_{ho}} \right)
\; .
\end{eqnarray}
For magnetic fields between $B_0$ and $B_{\rm res}$, the confinement thus stabilizes the molecules against dissociation that would occur in free space \cite{moritz:05}.

We measure $B_0=413.90$~G \cite{atom-molecule-paper:414G} and fit Eq.~(\ref{eq-B-res-shift}) to the data in Fig.~\ref{fig-density}(b) where we use $V_0$ as the only free fit parameter, entering the equation in $\omega_{ho}(V_0)$ and $a_{ho}(V_0)$. The fit yields a value of $V_0$ that is a factor of 1.20(2) larger than our independent calibration. This might mean that our independent calibration of $V_0$ is worse than we think it is, or else it means that other mechanisms have a significant effect. For example, the harmonic approximation used to derive Eq.~(\ref{eq-B-res-shift}) might be too inaccurate or the observed deviation might be explained as a differential ac-Stark shift between states $|\psi_a\rangle$ and $|\psi_m\rangle$ induced by the lattice light. The size of the differential ac-Stark shift would have to equal 2\% of the total ac-Stark shift.

In a final measurement, we study the dissociation of $|\psi_m\rangle$ into excited trap states of the entrance channel. To this end, we induce Rabi oscillations as in Fig.~\ref{fig-oscillation} and then stop the oscillations at a point where the molecule fraction is large, by jumping the magnetic field to a different value, typically far above the lowest oscillator state. We hold the field there for 250~$\mu$s and then switch it off.

\begin{figure}[t!]
\includegraphics[scale=.5]{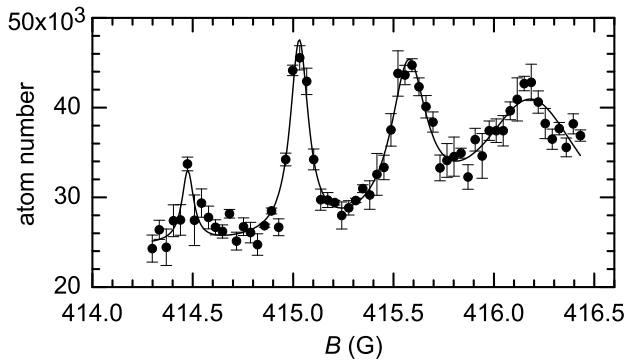}
\caption{\label{fig-band-structure}
Mapping the band structure using molecule dissociation. The separation of the peaks corresponds to $\sim 2\hbar \omega_{ho}$, i.e.\ the molecules dissociate into even bands. Dissociation is suppressed in the band gaps, showing the existence of confinement-induced band-gap molecules. The data were obtained at $V_0=25\,E_r$. The line is a guide to the eye.
}
\end{figure}

The observed number of entrance-channel atoms is shown in Fig.~\ref{fig-band-structure}. The peak near 414.5~G corresponds to the lowest oscillator state. The signal at this peak is fairly small because the dissociation pulse duration of $250~\mu$s happens to be close to a minimum of the Rabi oscillations, where only few molecules are dissociated. The peaks are approximately equidistant with a separation of $\sim 0.57$~G, corresponding to an energy difference of $\sim 2\pi\hbar\times 63$~kHz which is close to $2\hbar\omega_{ho}=2\pi\hbar\times 66$~kHz. Figure \ref{fig-band-structure} shows a suppression of dissociation into odd bands of the lattice as well as into band gaps. The suppression in the gaps demonstrates that confinement-induced molecules can be created not only below the lowest band \cite{moritz:05} but also in three-dimensional band gaps. We attribute the suppression of dissociation into odd bands to the fact that these bands have odd parity at quasi momentum zero. Hence, the $s$-wave state $|\psi_m\rangle$ cannot dissociate at quasi momentum zero. This measurement also profits from the reduced sensitivity to magnetic field noise due to the small value of $\Delta \mu$.

Even when avoiding eddy currents as above, we see no Rabi oscillations for dissociation into the second band near 415.0~G. We attribute this to single-atom tunneling, which has an amplitude of $J=2\pi\hbar\times 3.8$~kHz in the second band and leads to decoherence.

In conclusion, we observed large-amplitude atom molecule Rabi oscillations in an optical lattice using a Feshbach resonance. The dependence of the oscillations on magnetic field and lattice depth agrees well with a theoretical model. In addition, we observed confinement-induced band-gap molecules.

The oscillations observed here can be used for precisions measurements of atomic scattering properties that could be employed in sensitive tests for drifts of fundamental constants. In addition, confinement-induced molecules offer more general perspectives to manipulate the stability of molecules by structured environments. Finally, the production of a coherent atom-molecule superposition state with controllable amplitude and phase opens up new possibilities for quantum simulations.

We thank S. Kokkelmans for providing predictions for the parameters of the Feshbach resonance. We acknowledge financial support of the German Excellence Initiative via the program Nanosystems Initiative Munich.

\subsection{Appendix:\\ Matrix Element for Atom-Molecule Coupling}
Here, we derive the matrix element $H_{am}$ for atom-molecule coupling in a harmonic oscillator, given in Eq.~(1). In order to relate $H_{am}$ to parameters of the scattering problem, we first summarize some results of scattering theory. The asymptotic form of the scattering state in the absence of the molecular state is
\begin{eqnarray}
\label{eq-scattering}
\psi_{\bf k}({\bf r}) = C \left( e^{i{\bf k\cdot r}} + f_{bg} \frac{e^{ikr}}r \right)
\, ,
\end{eqnarray}
where $\bf r$ is the relativ coordinate, $\bf k$ is the wave vector in the relative coordinate, $f_{bg}=(e^{-2ika_{bg}}-1)/2ik$ is the background scattering amplitude and $C$ is a constant determined by normalization of the wave function. Obviously, $\psi_{\bf k}({\bf r})$ is proportional to $C$, so that
\begin{eqnarray}
\label{eq-alpha}
\langle\psi_m |H| \psi_{\bf k} \rangle = C \alpha
\, .
\end{eqnarray}
For small enough $k$, the parameter $\alpha$ becomes independent of ${\bf k}$ \cite{timmermans:99}. For the scattering state Eq.~(\ref{eq-scattering}), we use a large quantization volume $V$ and obtain $|C|^2=1/V$. We thus recover Eq.~(23) of Ref.~\cite{timmermans:99}. A combination of Eqs.~(25) and (42) of that reference yields
\begin{eqnarray}
|\alpha|^2 = \frac{4\pi\hbar^2 a_{bg}}m \;  \Delta\mu \Delta B
\, .
\end{eqnarray}
This relates $\alpha$ to known parameters of the Feshbach resonance. 

We now turn to the problem of two identical bosons in a harmonic oscillator. Our following treatment rests on the central assumption that there is a separation of length scales: the typical range of the interatomic interaction is much smaller than $a_{ho}$. Hence, $H_{am}$ can only be sensitive to the properties of the entrance-channel wave function at short radius. For the ground state of two interacting particles in the harmonic oscillator, the short-range behavior is \cite{busch:98}
\begin{eqnarray}
\label{eq-psi-ho}
\psi_a(r) = - \sqrt{\frac1{4\pi a_{ho}^2 \hbar\omega} \; \frac{\partial E}{\partial a_{bg}} }
\left[ -\frac{a_{bg}}r + 1 +O\left(\frac{r}{a_{ho}}\right) \right]
\end{eqnarray}
with the energy $E(a_{bg})$ given by the implicit equation \cite{busch:98}
\begin{eqnarray}
\frac{a_{ho}}{a_{bg}}
= \sqrt2 \; \frac{\Gamma\left(\frac34 -\frac E{2\hbar\omega}\right)}
{\Gamma\left(\frac14 -\frac E{2\hbar\omega}\right)}
\, ,
\end{eqnarray}
where $\Gamma$ is the Euler gamma function. A series expansion of this implicit equation yields for the ground state
\begin{eqnarray}
\frac{\partial E}{\partial a_{bg} }
= \sqrt{\frac 2\pi} \; \frac{\hbar\omega}{a_{ho}} \left[ 1+ 0.490 \, \frac{a_{bg}}{a_{ho}} + O\left(\frac{a_{bg}}{a_{ho}}\right)^2 \right]
\, .
\end{eqnarray}
In order to connect these results to the scattering problem, we consider Eq.~(\ref{eq-scattering}) in the limit $k\rightarrow 0$
\begin{eqnarray}
\label{eq-scattering-small-r}
\psi_{\bf k}(r)
&=&  C e^{-ika_{bg}} \left[1 +O(ka_{bg})^2\right] \nonumber \\
&& \times \left[ -\frac{a_{bg}}r + 1 + O(k r) + O(ka_{bg})^2 \right]
\! . \;\;
\end{eqnarray}
We note that the typical wave vector in the harmonic oscillator ground state is $k \sim 1/a_{ho}$, so that neglecting terms of order $O(ka_{bg})^2$ is consistent with our above series expansion that neglects terms of order $O(a_{bg}/a_{ho})^2$. Evidently, Eq.~(\ref{eq-scattering-small-r}) becomes identical to Eq.~(\ref{eq-psi-ho}) if we choose
\begin{eqnarray}
|C|^2= \left(\frac1{\sqrt{2\pi} a_{ho}}\right)^3 
\left[ 1+ 0.490 \, \frac{a_{bg}}{a_{ho}} + O\left(\frac{a_{bg}}{a_{ho}}\right)^2 \right]
\! .
\end{eqnarray}
As the coupling of an entrance-channel state to the molecular state is only sensitive to the wave function at short radius, and as the two wave functions at short radius Eqs.~(\ref{eq-psi-ho}) and (\ref{eq-scattering-small-r}) are identical, the coupling for the two states is identical. Hence, Eq.~(\ref{eq-alpha}) is again applicable and we obtain
\begin{eqnarray}
\lefteqn{|H_{am}|^2 =} && \nonumber \\
&=& \frac{4\pi \hbar^2 a_{bg} \Delta\mu \Delta B}{m \left( \sqrt{2\pi} \; a_{ho}\right)^3}
\left[ 1+ 0.490 \frac{a_{bg}}{a_{ho}} + O \left( \frac{a_{bg}}{a_{ho}} \right)^2 \right]
\! . \;\;\;
\end{eqnarray}


\begin{thebibliography}{10}
\bibitem{atom-molecule-paper:julienne}
Equation (22) in Ref.~\cite{julienne:04} is a factor of $(3/\pi)^{1/4}\sim
  0.99$ smaller and neglects the term $a_{bg}/a_{ho}$.
\bibitem{atom-molecule-paper:eddy}
The signal in Fig.~1 has a minimum at $t=0$ because for this data set, the
  first magnetic-field step ends 50~mG above the Feshbach resonance. Hence, the
  eddy currents effectively lead to an adiabatic ramp across the Feshbach
  resonance and associate molecules, as in Ref.~\cite{mark:05}.
\bibitem{atom-molecule-paper:414G}
This value is 0.44~G smaller than in Ref.~\cite{marte:02}. We attribute this to
  the fact that the loss feature observed in Ref.~\cite{marte:02} at this
  particular resonance was extraordinarily broad, making an accurate
  determination of the resonance position in Ref.~\cite{marte:02} very
  difficult.
\end{thebibliography}
\end{document}